\documentclass[
 reprint,
 amsmath,amssymb,
 aps,
prl,
]{revtex4-2}

\usepackage{graphicx}
\usepackage{siunitx}
\usepackage{hyperref}
\usepackage{booktabs}
\usepackage{tabularx}

\newcommand*{\figref}[2][]{%
	\hyperref[{fig:#2}]{%
		Fig.~\ref*{fig:#2}%
		\ifx\\#1\\%
		\else
		(#1)%
		\fi
	}%
}

\newcommand*{\figsref}[2][]{%
	\hyperref[{fig:#2}]{%
		Figs.~\ref*{fig:#2}%
		\ifx\\#1\\%
		\else
		(#1)%
		\fi
	}%
}

\newcommand*{\figureref}[2][]{%
	\hyperref[{fig:#2}]{%
		Figure~\ref*{fig:#2}%
		\ifx\\#1\\%
		\else
		(#1)%
		\fi
	}%
}

\newcommand*{\figuresref}[2][]{%
	\hyperref[{fig:#2}]{%
		Figures~\ref*{fig:#2}%
		\ifx\\#1\\%
		\else
		(#1)%
		\fi
	}%
}

\newcommand*{\fref}[2][]{%
	\hyperref[{fig:#2}]{%
		\ref*{fig:#2}%
		\ifx\\#1\\%
		\else
		(#1)%
		\fi
	}%
}

\newcommand*{\tabref}[2][]{%
	\hyperref[{tab:#2}]{%
		Tab.~\ref*{tab:#2}%
		\ifx\\#1\\%
		\else
		(#1)%
		\fi
	}%
}

\newcommand*{\tableref}[2][]{%
	\hyperref[{tab:#2}]{%
		Table~\ref*{tab:#2}%
		\ifx\\#1\\%
		\else
		(#1)%
		\fi
	}%
}

\begin{document}

\title{Scalable Multilayer Architecture of Assembled Single-Atom Qubit Arrays\\
	 in a Three-Dimensional Talbot Tweezer Lattice}
\author{Malte~Schlosser}
\email{apqpub@physik.tu-darmstadt.de}
\author{Sascha~Tichelmann}
\author{Dominik~Sch\"affner}
\author{Daniel~Ohl~de~Mello}
\author{Moritz~Hambach}
\affiliation{Technische Universit\"{a}t Darmstadt, Institut f\"{u}r Angewandte Physik, Schlossgartenstra\ss e 7, 64289 Darmstadt, Germany}
\author{Jan~Sch\"utz}
\altaffiliation{Present address: Fraunhofer Institute for Physical Measurement Techniques IPM, Georges-K\"ohler-Allee 301, 79110 Freiburg, Germany.}
\author{Gerhard~Birkl~\hspace{1mm}}
\homepage{https://www.iap.tu-darmstadt.de/apq}
\affiliation{Technische Universit\"{a}t Darmstadt, Institut f\"{u}r Angewandte Physik, Schlossgartenstra\ss e 7, 64289 Darmstadt, Germany}

\date{\today}

\begin{abstract}
	We report on the realization of a novel platform for the creation of large-scale 3D multilayer configurations of planar arrays of individual neutral-atom qubits:~a microlens-generated Talbot tweezer lattice that extends 2D tweezer arrays to the third dimension at no additional costs. We demonstrate the trapping and imaging of rubidium atoms in integer and fractional Talbot planes and the assembly of defect-free atom arrays in different layers. The Talbot self-imaging effect for microlens arrays constitutes a structurally robust and wavelength-universal method for the realization of 3D atom arrays with beneficial scaling properties. With more than 750 qubit sites per 2D layer, these scaling properties imply that \num{10000} qubit sites
	are already accessible in 3D in our current implementation. The trap topology and functionality are configurable in the micrometer regime. We use this to generate interleaved lattices with dynamic position control and parallelized sublattice addressing of spin states for immediate application in quantum science and technology.\\	
	
	\doi{10.1103/PhysRevLett.130.180601}
	%
\end{abstract}

\maketitle
Rapid advances in experimental platforms for quantum science and technology propel progress in a wide range of forefront research: exploiting the laws of quantum mechanics, experiments promote quantum chemistry \cite{Julienne2012}, quantum interferometry \cite{Cronin2009}, quantum metrology \cite{Ludlow2015,Degen2017}, and quantum information science \cite{Georgescu2014,Alexeev2021,Altman2021,Bharti2022}.
One of the signature architectures are arrays of neutral atoms in optical tweezers \cite{Kaufman2021}, which constitute a versatile experimental platform with comprehensive coherent control of quantum states and tunable interactions at the single quantum level. Their spatially resolved trapping sites are adaptable to hold individual atoms \cite{Ott2016,Brown2019,Jenkins2022} and can be composed of multiple colors of laser light \cite{Schlosser2019,Barredo2020,Zhang2022,Graham2022}. Straightforward dynamic position control facilitates the coherent transport of quantum states \cite{Beugnon2007,Lengwenus2010,Kaufman2015,Dordevic2021,Bluvstein2022} and the seminal assembly of defect-free subarrays \cite{Adams2019,Browaeys2020,Henriet2020,Morgado2021,Kaufman2021}. Thus, arrays of individually controlled microtraps lend themselves to the bottom-up engineering of quantum systems with configurable tunnel coupling \cite{Murmann2015,Kaufman2015,Sturm2017,Kaufman2021,Spar2022} and on-demand Rydberg interactions \cite{Saffman2016,Browaeys2016,Adams2019,Browaeys2020,Henriet2020,Morgado2021,Ebadi2021,Scholl2021,Kaufman2021,Ma2022,Graham2022,Chew2022}.\\
%
\begin{figure}
	\includegraphics[width=1\linewidth]{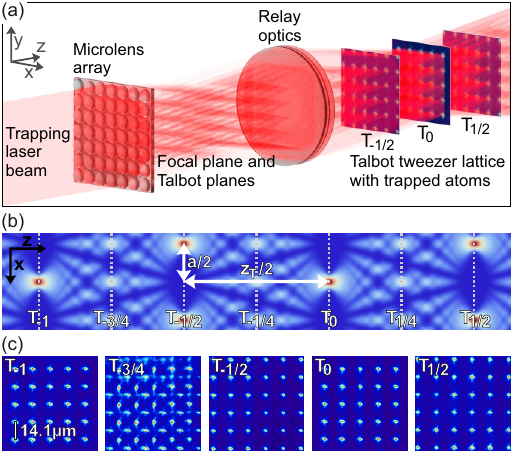}
	\caption{Microlens-array generated Talbot tweezer lattice. 
		(a) Schematic of the experimental setup. The MLA generates a 2D periodic spot pattern in its focal plane that is reproduced in Talbot planes. After reimaging and demagnification, these recurring self-images form a regular 3D multilayer lattice of 2D periodic optical microtrap arrays (for clarity, only three planes are drawn).
		(b) Detail of a 2D projection of the calculated intensity distribution (color coded) associated with a Talbot tweezer lattice created by an array of Gaussian beams. The principal planes $T_\mathrm{0},T_\mathrm{\pm1/2},T_\mathrm{\pm1}, ...$ are separated by ${z_T/2}$ and show a trap pitch identical to the one of the focal plane $T_\mathrm{0}$. Parameters are according to the experiment: ${\lambda=\SI{798.6}{\nano\meter}}$, ${a=\SI{14.1(4)}{\micro\meter}}$, ${z_T/2=\SI{248(14)}{\micro\meter}}$, ${w_0=\SI{1.45(10)}{\micro\meter}}$.
		(c) Averaged fluorescence images of single-atom arrays in integer and fractional Talbot planes (see text for details).
	}\label{fig:talbot}
\end{figure}%
%
For future progress, scalability is pivotal. In this Letter, we resolve this issue with a configurable large-scale implementation using a novel platform for quantum simulation and computing: we harness the Talbot effect \cite{Wen2013,Besold1997a,Mennerat1998,Ovchinnikov2006} to extend a microlens-array (MLA) generated 2D periodic configuration of single-atom microtraps to a multilayer architecture in the third dimension (\figref[]{talbot}). This Talbot tweezer lattice unlocks a new class of 3D atomic arrays with excellent scalability as it is built in parallel from a microfabricated source element with tens of thousands of lenslets \cite{OhldeMello2019,Voelkel2012} and the self-imaging effect that extends the 2D tweezer array to 3D at no additional costs.\\
\figureref[a]{talbot} visualizes the experimental platform: the source element is a quadratic-grid MLA that is illuminated by a trapping laser beam with a Gaussian profile and produces a 2D periodic spot pattern in its focal plane. This is the origin of a 3D Talbot lattice, which exhibits integer and fractional self-images of the generating pattern in the axial dimension, referred to as Talbot planes \cite{supplement}\nocite{Zappe2012,Gonzalez2022,Schaffner2020,Kirner2019,Schlosser2011,Pause2023,Gillen2006,Li1981,Ruffieux2006,Winthrop1965,Besold1997b,Patorski1989}. The separation of integer self-images is given by the Talbot length ${z_T=2a^2/\lambda}$, which is a quadratic function of the spatial period ${a}$ of the generating pattern. In our setup, the MLA's focal plane with a pitch identical to the pitch of the MLA is reimaged and demagnified by the use of relay optics, which renders the trap size $w_0$, trap pitch $a$, and the associated Talbot length configurable. Individual planes $T_i$ are labeled according to their axial position normalized to ${z_T}$ with $T_0$ being the reimaged focal plane. An important modification due to reimaging is the appearance of negative indexed Talbot planes before the focal plane $T_0$ that complete the symmetric 3D structure of the resulting Talbot tweezer lattice. Principal planes, such as $T_{0},T_{\pm1/2},T_{\pm1}, ...$, show the same in-plane trapping geometry and pitch as the focal plane and are separated by ${z_T/2}$. The planar arrays in consecutive principal planes are shifted by ${a/2}$ in the x and y direction. 
A typical MLA, as used in the realization shown in \figref[]{talbot}, consists of $166{\times}166 = \num{27556}$ refractive lenslets with a pitch of \SI{30.0(3)}{\micro\meter} (see figure caption for experimental parameters). \figureref[b]{talbot} shows a 2D projection of the simulated Talbot tweezer lattice in the relevant region along the optical axis.\\
During a sequence of optical molasses, laser cooled $^{85}$Rb atoms are prepared in the lattice traps with typical depths on the order of ${k_B\times\SI{1}{\milli\kelvin}}$. Because of the collisional blockade effect \cite{Ott2016,Brown2019}, each site is occupied by one atom at most. For the preparation and imaging of a 2D array of atoms in a selected Talbot plane, the imaging system is focused to the selected plane and the atoms in all other planes are removed with a resonant blow-away laser beam. \figureref[c]{talbot} exemplarily shows the averaged fluorescence images of individual atoms stored in the traps of the reimaged focal plane $T_0$ and of the Talbot planes $T_{-1}$, $T_{-3/4}$, and $T_{\pm1/2}$. The trap positions are stable and optically resolved which reveals the underlying periodic geometry. For the fractional plane $T_{-3/4}$ an increase in the number of trapping sites is observed.
As expected, the trap pitch is reduced by a factor of 2, but the number of trapping sites with visible atomic fluorescence differs from a simple model: the signal is weaker and even vanishes for a subclass of sites at regular positions. We attribute the observed behavior to a modulation of the peak intensities within the plane resulting, for certain positions, in a reduction of the trap depth to the limit where atoms are no longer retained.
This effect arises from interference of light transmitted by the lenslets and light transmitted through the interstices of the MLA as well as from components of the light field that are not included in a simple paraxial description and originate in the spherical shape and limited aperture of the lenslets \cite{Besold1997a,Kim2015}.
We do not observe these effects in the principal Talbot planes and thus limit our experiments to atoms within these principal planes.
\\
%
\begin{figure}[t]
	\includegraphics[width=\linewidth]{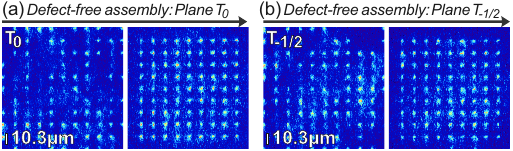}
	\caption{
		In-plane assembly of defect-free target struc\-tu\-res of individual atoms using an auxiliary transport tweezer. The reimaged focal plane $T_0$ (a) and the Talbot plane $T_{-1/2}$ (b) are shown (atom arrays are centered in the images due to cropping). Pa\-ra\-me\-ters: ${\lambda=\SI{797.3}{\nano\meter}}$, ${a=\SI{10.3(3)}{\micro\meter}}$, ${z_T/2=\SI{133(8)}{\micro\meter}}$, ${w_0=\SI{1.45(10)}{\micro\meter}}$, transport tweezer size (waist) \SI{2.0(1)}{\micro\meter}.
		The \textit{in situ} atomic fluorescence of individual atoms at resolved trapping positions reveals the initial atom distribution [left in (a),(b)] and the successful creation of a ${9 \times 9}$ cluster via atom-by-atom assembly in both Talbot planes [right in (a),(b)].
		The assembly process involves multiple cycles of atom rearrangement and position detection.
	}\label{fig:sa}
\end{figure}%
%
The stochastic nature of the single-atom loading process is reflected by the random distribution of stored atoms within the 3D lattice, as pictured in \figsref[a]{sa} (left) and \fref[b]{sa} (left) for the tweezer arrays of planes ${T_0}$ and ${T_{-1/2}}$. We observe a maximum probability of 0.6 to load an atom for central sites. This necessitates the incorporation of techniques for individual atom transport \cite{Beugnon2007,Lengwenus2010,Kaufman2015,Dordevic2021,Bluvstein2022} and deterministic target pattern assembly \cite{Bluvstein2022,Adams2019,Browaeys2020,Henriet2020,Morgado2021,Kaufman2021}, as a broad range of applications prerequires defect-free atom arrays and the ability to mend atom loss during operation. For this purpose, we use a superposed movable optical tweezer. This technique lends itself to 3D layered configurations of tweezer arrays \cite{Barredo2018} and is applicable to any Talbot plane in a straightforward fashion as demonstrated in this Letter.
In our setup, the transport tweezer is aligned with the imaging system and therefore automatically focused to the selected plane. The principle of operation is detailed in \cite{OhldeMello2019,Schlosser2020}, where it has been applied to the rearrangement of atoms in the plane ${T_0}$ so far. Starting from the initial atom distribution, which serves as a reservoir, a sequence of transport operations is performed in order to reach complete filling of a predefined target structure.
We define a ${9 \times 9}$ atom cluster as target structure and operate on a reservoir grid of ${19 \times 19}$ sites. On average, we initially prepare 191(17) atoms.
\figuresref[a]{sa} and \fref[b]{sa} show fluorescence images of the atom pattern before and after successful assembly in the planes ${T_0}$ and ${T_{-1/2}}$. After rearrangement, we observe an average filling fraction on the order of \num{0.9} and a typical success rate for a defect-free cluster of \SI{12}{\%}. This demonstrates that defect-free atom assembly is possible in all principle planes of the Talbot tweezer lattice.\\%
%
\begin{figure}[t]
	\includegraphics[width=\linewidth]{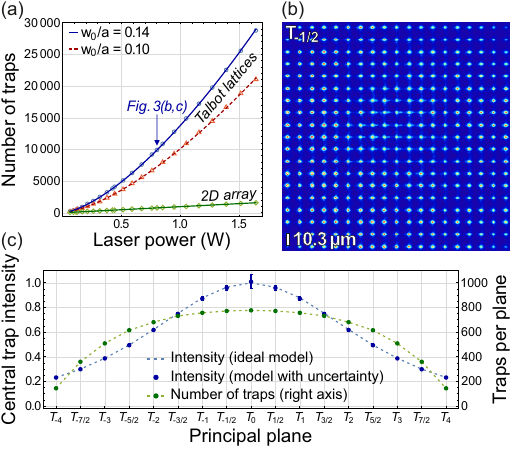}
	\caption{
		Scaling properties of the Talbot tweezer lattice and large-scale implementation.
		(a) The integral number of traps of all principal planes in the 3D lattice shows a scaling of $N_\mathrm{int}\propto P^{3/2}$ as function of the trap laser power. The scaling is plotted for the two experimental values of ${w_0/a}=\SI{0.10}{}$ [red dashed, corresponding to \figsref[c]{talbot} and \fref[]{interleaved}] and ${w_0/a}=\SI{0.14}{}$ [blue, corresponding to \figsref[]{sa} and \fref[b]{scaling}]. In one plane (2D array), the number of traps is directly proportional to the laser power (green, linear scaling).
		(b) Experimental implementation of a 3D tweezer lattice with \num{e4} traps [see indication in (a) and text for details]. The image shows a ${19 \times 19}$ detail of single atoms stored in the plane ${T_{-1/2}}$ (averaged fluorescence).
		(c) Intensity and trap number variation along the Talbot tweezer lattice of \figref[b]{scaling}. The intensity of the central trap normalized to $T_0$ is given for the ideal model (dashed blue line) and the model including uncertainties (blue dots). Green dots depict the in-plane number of traps (right axis).		
	}\label{fig:scaling}
\end{figure}%
%
As a consequence, the Talbot tweezer lattice exhibits superior scalability by extending the assembled atom configurations from 2D to 3D. The 3D scaling properties as a function of laser power $P$, i.e.,~the number of trapping sites per principal plane multiplied by the number of planes with sufficient trap depth, are governed by the following argument:
within the 2D focal plane, the number of traps is identical to the number of illuminated lenslets which is proportional the square of the trap beam radius at the position of the MLA. The Talbot effect gives rise to an additional linear increase in the number of suitable principal Talbot planes and thus in the total number of traps as a function of the trap beam radius. In total, this leads to a cubic scaling of the integral trap number $N_\mathrm{int}$ in all principal planes of the 3D tweezer lattice.
On the other hand, the required laser power increases linearly with the number of illuminated lenslets and quadratically with the trap beam radius at the MLA. No extra laser power is required for trapping of atoms in the additional Talbot planes. Extending to the third dimension comes for free!
Furthermore, $N_\mathrm{int}$ is proportional to the ratio $w_0/a$ of trap size and trap pitch. An increased spot size reduces the divergence of the propagating laser beams and accordingly reduces walk-off effects in the Talbot planes. A similar effect results from a reduction of the trap pitch, as this contracts the Talbot lattice.
As usual for optical traps, the required power scales quadratically with the inverse trap size: $P\propto w_0^{-2}$. Hence, we obtain the overall scaling $N_\mathrm{int}\propto P^{3/2}/(a w_0)$.\\
\figureref[]{scaling} depicts these Talbot-specific scaling properties alongside an experimental implementation of a single-atom Talbot tweezer lattice optimized for a large number of traps. In \figref[a]{scaling}, the integral number of traps $N_\mathrm{int}$ of all principal planes is given as a function of the trap laser power on the MLA. To calculate this number, the Talbot tweezer lattice is simulated by propagating an array of Gaussian beams \cite{supplement}. Traps shallower than one-fifth of the normalized trap depth are neglected, which is in agreement with the experimental observations. For comparison, the green curve visualizes the linear scaling of the number of traps in one Talbot plane as a function of laser power.
Obviously the scaling of the integral number of traps surpasses the scaling of the number of traps in one plane:~the traps in the additional Talbot planes come for free.
The image of \figref[b]{scaling} shows averaged fluorescence of the ${T_{-1/2}}$ plane of our largest implemented Talbot tweezer lattice. In this realization, the trapping laser was set to a wavelength of \SI{796.3}{\nano\meter} and a radius (waist) of \num{17.5} lens pitches. A total power of ${P}=\SI{0.803}{W}$ was illuminating the lens array. The ratio of trap size and trap pitch was ${w_0/a}=\SI{0.14}{}$.
These parameters result in 17 relevant principal planes originating from an array of 777 tweezers of sufficient depth in the plane ${T_{0}}$.
The variation along the principal planes is shown in \figref[c]{scaling}. For the central tweezer intensity, a comparison of the ideal model (dashed blue line) to the results obtained when accounting for uncertainties of the MLA parameters (blue dots) is given. As evident, the two models show only minor deviations. Furthermore, the self-imaging process results in a healing of perturbations for Talbot planes apart from $T_0$.
A detailed analysis of the properties of the MLAs used in this work and of the resulting Talbot tweezer lattices, as well as the impact of light field imperfections and production tolerances of MLAs are given in the Supplementary Material \cite{supplement}.
The in-plane trap count is depicted in green (dots, right axis).
In total, we calculate a number of \num{e4} traps of sufficient depth for this Talbot tweezer lattice, as presented in \figref[a]{scaling}.
Within the readily accessible limits of titanium-sapphire laser technology, the power illuminating the MLA can be scaled by at least a factor of 5 beyond the given power. This increases the integral number of traps $N_\mathrm{int}$ in the 3D lattice by a factor of 11, pushing the number of attainable single atom qubit sites beyond \num{e5} based on the parameters of \figref[b]{scaling}. No specific measures for reducing the spectral width beyond the typical parameters of commercial titanium-sapphire lasers are required due to the small variation of the lengths of the beam paths contributing to each Talbot spot \cite{Huft2022}.\\
In order to further explore the potential of Talbot tweezer lattices for quantum computation and simulation, we combine the scalability of our platform with the ability to parallelize quantum operations using interleaved Talbot lattices. This technique enables the dynamic modification of configurations in interleaved 2D subarrays and furthermore the realization of 3D layered geometries with layer separations not limited to the Talbot length. Thereby the class of accessible geometries is significantly expanded to elaborate arrangements that incorporate parallelized transport \cite{Lengwenus2010,Brown2019,Bluvstein2022}, enable multispecies trapping \cite{Saffman2016,Kaufman2021,Zhang2022,Singh2022b}, can be configured to implement atom subgroups and ancilla atoms \cite{Weimer2010,Morgado2021,Sheng2022,Singh2022b}, and give access for applying inhomogeneous interaction potentials \cite{Leseleuc2019}. Versatile interleaved and layered Talbot tweezer lattices can be implemented by employing multiple MLAs \cite{Lengwenus2010, Schlosser2012} as well as by the illumination of a single MLA with multiple laser beams. The latter approach, as presented here, results in axially superposed lattice planes if the interleaved tweezer lattices are of near-identical wavelength. Lateral displacement of the lattice sites is achieved through a modification of the incident angle on the MLA. While the displacement is a linear function of the MLA's focal length for the plane $T_0$, for all other Talbot planes one has to consider the effective focal length, which is enlarged by the absolute value of their axial separation from the focal plane before reimaging. Hence, working in a Talbot plane significantly reduces the required angular displacement for achieving a sufficient lateral displacement. This allows one to dynamically compose tweezer arrays of versatile geometries with flexible site separation from superposed movable subarrays that can be controlled independently.\\
%
\begin{figure}
	\includegraphics[width=1\linewidth]{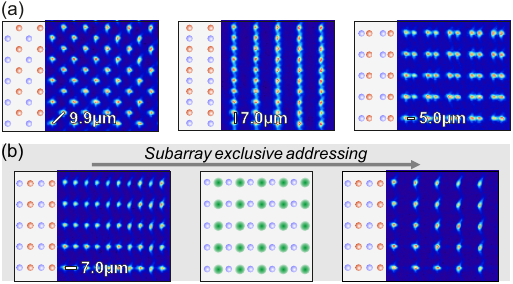}
	\caption{Interleaved configurations of two Talbot tweezer lattices. All images show averaged fluorescence of the ${T_{1/2}}$ plane and the geometries are visualized by red and blue dots depicting the sites of the two lattices.
		(a) Single atom arrays with tunable atom separation. The separation of neighboring traps is (from left to right): \SI{9.9(3)}{\micro\meter}, \SI{7.0(2)}{\micro\meter}, and \SI{5.0(2)}{\micro\meter}.
		(b) Sublattice-exclusive addressing (green, middle) is used to convert a spin-polarized array (${F=2}$, left) to a configuration of columnwise antiferromagnetic ordering via preparation of alternating spin states (${F=2}$ and ${F=3}$, right). State-selective detection records fluorescence for atoms in ${F=2}$.}\label{fig:interleaved}
\end{figure}%
%
\figureref[]{interleaved} shows configurations of two tweezer subarrays in the Talbot plane ${T_{1/2}}$ which are created from two interleaved lattices. These are implemented by extending the setup of \figref[]{talbot} by a second trapping laser beam illuminating the same MLA under a variable angle.
The two beams of equal size are set to a frequency difference of \SI{30.0}{\mega\hertz} to avoid interference effects. An enclosed angle of \SI{12.3}{\milli\radian} introduces an offset of $a/2$ between the respective subarrays. \figureref[a]{interleaved} (left) shows a uniform arrangement that reduces the trap separation by a factor of ${\sqrt{2}}$, with each site of one respective subarray (red) being equidistant to four sites of the other one (blue).
Furthermore, the trap positions can be tuned to create atom chains and to favor pairs of atoms as presented in  \figref[a]{interleaved} (middle and right). In the current implementation, traps stay independent for a minimum separation of \SI{3}{\micro\meter}, which is consistent with the trap waist.
Considering applications of interacting atoms (e.g.,~via Rydberg interactions) the latter two asymmetric configurations allow for the implementation of selectable interaction geometries with interactions taking place, e.g.,~only along the columns as in \figref[a]{interleaved} (middle) or only between multiple realizations of atom pairs as in \figref[a]{interleaved} (right).\\
Laser-induced state manipulation and atom-atom coupling also profit from our approach: the sites of each sublattice remain independently laser-addressable in a parallelized fashion through the MLA, which focuses the addressing light onto the stored atoms and thereby amplifies state manipulation and atom-light coupling. If the addressing beam is set to spatial overlap with the trapping light before illuminating the MLA, it is auto-aligned with the traps while the spacing of interleaved tweezer arrays in a selected Talbot plane can be arbitrarily set within the limits given by trap pitch and trap waist.\\
In the experiments of \figref[b]{interleaved} we demonstrate parallelized laser addressing in sublattices with the transfer of the spin state of single atoms stored in tweezer subarrays with a spacing of \SI{7.0(2)}{\micro\meter}. First, all atoms in the full set of traps are initialized to the ${F=2}$ state of the $^{85}$Rb ground-state hyperfine manifold and state-selectively detected afterwards by removing the population of the ${F=3}$ state prior to detection (left). The results displayed on the right account for a modified experimental sequence where laser addressing transfers the atoms to the ${F=3}$ state in a sublattice-exclusive fashion after initial state preparation, resulting in vanishing fluorescence at the addressed sites. This presents atom chains exhibiting antiferromagnetic ordering along the horizontal dimension in \figref[b]{interleaved}. 
The addressing light at \SI{795}{\nano\meter} is brought to spatial overlap with the respective sublattice by superposing it with the respective trapping laser beam using a single mode optical fiber. For the center trap, the measured occupation probability after initial state preparation is \num{0.47(2)}. This value is reduced to \num{0.01(1)} when including the subarray-exclusive state transfer before the state-selective detection. For the unaddressed subarray, we observe a reduction on the order of \SI{2}{\%}. We contribute this crosstalk to chromatic aberrations causing imperfect overlap of the addressing light and addressed sites as well as residual scattering in the unaddressed sublattice.\\
In summary, we have created Talbot tweezer lattices unfolding multisite 3D atom configurations with customizable structure sizes in the micrometer regime. With the presented work, microlens-generated tweezer arrays enter a regime that has formerly been accessible only by phase-sensitive standing wave configurations \cite{Gross2017,Kumar2018}, yet they retain the simplicity and robustness of the MLA-based optical setup as well as their inherent features of coherent quantum state transport \cite{Lengwenus2010} and single-site addressability for qubit and spin-state manipulation \cite{Kruse2010}.
The straightforward assembly of defect-free atom arrays facilitates bottom-up engineering of quantum systems in 2D and 3D geometries. In addition, atom separations can be dynamically adapted in 3D using layered interleaved configurations, while sublattice-exclusive addressability extends techniques for the initialization and readout of quantum states and allows for the parallelization of operations.
The implemented Talbot tweezer lattices provide several thousand of single-addressable atomic qubits with separations well within the limits for coherent Rydberg-mediated interactions \cite{Schlosser2020} with direct implications for neutral atom quantum information science.
Extending the reported results within the limits of available technology, Talbot tweezer lattices in the regime of \num{100000} quantum systems are accessible already to date.
\begin{acknowledgments}
	We acknowledge financial support by the Deutsche Forschungsgemeinschaft (DFG) [Grants No.~\mbox{BI\,\,647/6-1} and No.~\mbox{BI\,\,647/6-2}, Priority Program SPP\,\,1929 (GiRyd)] and by the Federal Ministry of Education and Research (BMBF) [Grant No.\,\,13N15981]. We thank L.~Pause, T.~Preuschoff, M.R.~Sturm, and R.~Walser for insightful discussions.
\end{acknowledgments}
\bibliography{multilayer}
\onecolumngrid
\renewcommand{\thetable}{S\arabic{table}}
\setcounter{table}{0}  
\renewcommand{\thefigure}{S\arabic{figure}}
\setcounter{figure}{0}
\renewcommand{\theequation}{S\arabic{equation}}
\setcounter{equation}{0}
~\\
\begin{large}\begin{center}
		\textbf{Supplementary Information:\\
			Scalable Multilayer Architecture of Assembled Single-Atom Qubit Arrays\\
			in a Three-Dimensional Talbot Tweezer Lattice}
\end{center}\end{large}

This document provides supplementary information on the characteristics of microlens arrays, the simulation of the generated light fields, and the requirements on the incident trapping laser light for the implementation of a Talbot tweezer lattice. In the second part, special emphasis is given to the effects of imperfections of microlens arrays resulting from production tolerances and the mitigation of perturbations induced by the self-imaging process.
\section{Microlens Arrays (MLA)}
The micro-fabrication of optical elements builds on a rich technological foundation that facilitates the production of complex refractive and diffractive structures for a variety of optical materials such as polymers and glasses \cite{Zappe2012,Voelkel2012,Gonzalez2022}. While direct laser writing has proven proficient for free-form micro-optics, multi-element systems and rapid prototyping \cite{Gonzalez2022,Schaffner2020}, lithographic techniques adapted from semiconductor processing enable the wafer-level batch fabrication of micro-optical elements with nanoscale precision on planar substrates \cite{Zappe2012,Voelkel2012,Kirner2019}. MLAs covering an extensive range of parameters are commercially available.\\
Focusing on regular-grid, two-dimensional arrays of converging lenses, as required for the implementation of an atom-optical  Talbot tweezer lattice, hexagonal- and quadratic-grid MLAs have been used in atom-optic experiments \cite{Schlosser2011,Schaffner2020,Pause2023}. Lenslets are either contiguous or of circular footprint. Chrome apertures and standard anti-reflection coatings can be applied. Typical numerical apertures are on the order of \numrange{0.01}{0.2}{} and diameter and pitch (i.e.,~grid spacing) are in the range of \SIrange{10}{200}{\micro\meter}. A quadratic grid pattern of \SI{100}{\micro\meter} pitch results in \num{3} million lenslets on an \SI{8}{inch} wafer.
\subsection{Optical Properties of MLAs used for this work}
\begin{table}[b]
	\begin{center}
		\begin{tabular}{p{25mm}p{14mm}p{23mm}p{16mm}p{19mm}p{12mm}p{18mm}p{16mm}p{12mm}}
			\toprule\addlinespace[6pt]
			\multicolumn{9}{c}{Parameters: Microlens arrays}\\\addlinespace[2pt]
			\cmidrule(lr){1-9}\\\addlinespace[-8pt]
			Parameter~set&Lenslet xy-pitch&Approx.\,\,lenslet\newline diameter\,\,($d$)&Radius\,\,of curvature\newline (ROC)&Geometrical\newline focal\,\,length ($f_g$)&Fresnel number (FN)&Effective\newline focal\,\,length ($f_e$)&Effective\newline numerical aperture&Talbot length ($z_T$)\\
			\midrule
			MLA1&\SI{30}{\micro\meter}&\SI{26.5}{\micro\meter}&\SI{42}{\micro\meter}&\SI{93}{\micro\meter}&\num{2.4}&\SI{76}{\micro\meter}&\num{0.17}&\SI{2.3}{\milli\meter}\\
			MLA2~\textit{(Fig.\,S1)}&\SI{110}{\micro\meter}&\SI{106}{\micro\meter}&\SI{860}{\micro\meter}&\SI{1897}{\micro\meter}&\num{1.9}&\SI{1403}{\micro\meter}&\num{0.04}&\SI{30.4}{\milli\meter}\\		
			\bottomrule
		\end{tabular}
	\end{center}
	\caption{MLA parameters for a wavelength of $\lambda=\SI{797}{\nano\meter}$. The refractive index $n(\lambda=\SI{797}{\nano\meter})$ of fused silica is \num{1.4534}. The MLAs' Talbot lengths are given by ${z_T=2 \text{(xy-pitch)}^2/\lambda}$. Uncertainties of relevant parameters are listed in \tabref{MLAtab3}.}
	\label{tab:MLAtab1}
\end{table}%
Both MLAs used for this work have been fabricated utilizing photolithography and reactive ion etching on fused silica. This material withstands cw laser powers of tens of kilowatts per centimeter (linear power density) that therefore effectively imposes no limit on atom tweezer experiments. The MLAs are of quadratic grid with circular lenslets of spherical profile. Anti-reflection coating has been applied for the relevant wavelengths (residual reflectance < \SI{0.5}{\percent}). Relevant parameters are listed in \tabref{MLAtab1}. The lenslet pitch, approximate diameter and radius of curvature of the spherical profile are given by the manufacturer. The remaining parameters have been calculated on this basis. The listed values are in agreement with simulations and experimental findings.\\
The geometrical focal length $f_g$ of a spherical lens relates to its radius of curvature by $f_g=\text{ROC}/(n(\lambda) - 1)$, the  numerical aperture can be calculated from $f_g$ with the lenslet diameter $d$ according to $\text{NA}\approx d/(2 f_g)$ and the Fresnel number is given by $\text{FN}=d^2/(4 \lambda f_g)$.\\
Special attention has to be taken for FN on the order of \num{5} and below, where diffraction at the lens aperture significantly modifies the focused beam. In the limiting case of a circular aperture with
\nocite{Gillen2006,Li1981,Ruffieux2006,Winthrop1965,Besold1997b,Patorski1989,Kim2015,Barredo2018,Schlosser2020,Huft2022,Singh2022b,Weimer2010,Sheng2022,Leseleuc2019,Schlosser2012,Gross2017,Kumar2018,Kruse2010}
vanishing refractive power \cite{Gillen2006,Huft2022}, the light field behind the aperture shows an intensity maximum at a distance of $d^2/(4 \lambda)$. This effect reduces the effective focal length $f_e$ and induces a relative focal shift for finite values of ROC that can be approximated by $\Delta f/f_g\approx(1+\pi^2 \text{FN}^2/12)^{-1}$ giving $f_e\approx f_g (1-\Delta f/f_g)$ \cite{Li1981,Ruffieux2006}. 
\subsection{MLA light fields}
\begin{figure}[t]
	\centering 
	\includegraphics[width=0.67\linewidth]{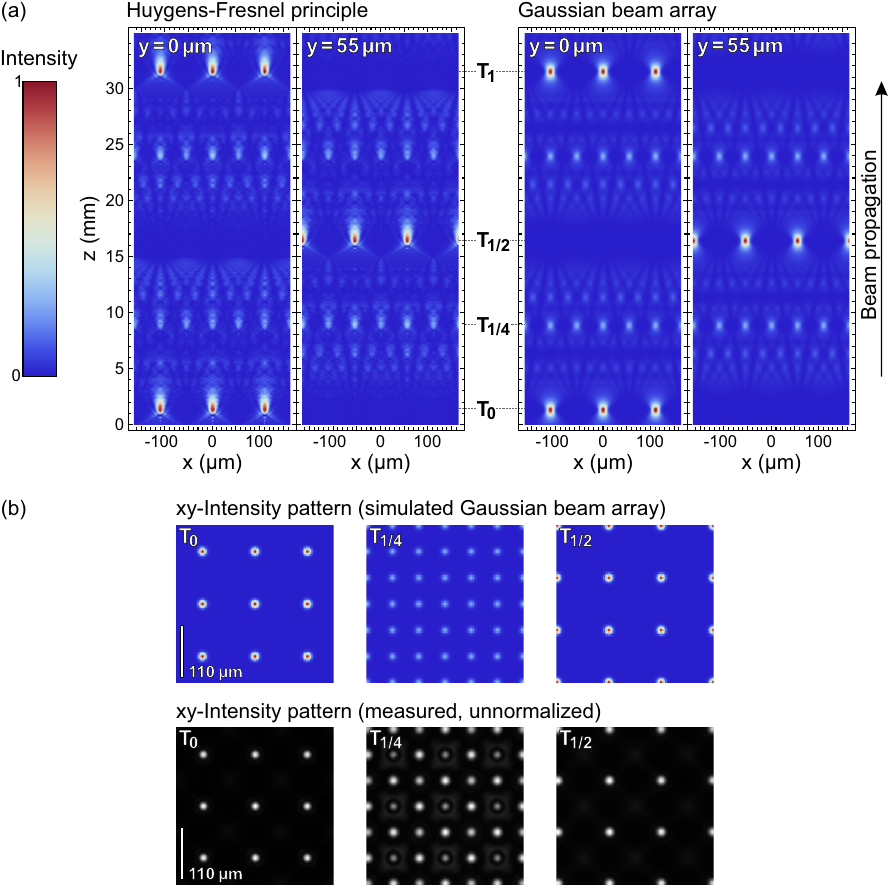} 
	\caption{Simulated and measured intensity patterns of the light field generated by MLA2 (see \tabref{MLAtab1}). The light propagates along the z-direction. (a) xz-sections of numerical data computed on basis of the Huygens-Fresnel principle (left) and a propagated Gaussian beam array (right) for two planes with $y=\SI{0}{\micro\meter}$ and $y=a/2=\SI{55}{\micro\meter}$, respectively. Plane $z=\num{0}$ corresponds to the position of the MLA (not shown). (b) xy-sections for three Talbot planes showing simulated data of the Gaussian beam array (top) and camera-recorded intensity patterns (bottom).}\label{fig:s1}
\end{figure}
The illumination of an MLA with a beam of coherent laser light whose lateral size is significantly larger than the lenslet pitch propagating along the z-direction produces a periodic spot pattern in the focal plane. The Talbot effect describes the appearance of regular self images of this spot array along the optical axis. For a quadratic grid array of spatial period (pitch) $a$, the axial position of the Talbot planes is given by \cite{Winthrop1965}:
\begin{align}\label{eqn:planes}
	z_{Q+M/N}= (Q+M/N)\cdot z_T\quad\text{with}\quad z_T= 2 a^2 /\lambda.
\end{align}
Here, the spatial period $a$ of the generating pattern corresponds to the lenslet xy-pitch, $z_T$ is the Talbot length and $Q$, $M$, $N$ are natural numbers for which the conditions $N > 0$, $0 \leq M < N$ apply. Integer Talbot planes ($M/N=0$) and the fractional Talbot planes with $M/N = 1/2$ fully recreate the generating pattern, while the planar arrays of the latter are shifted by $a/2$ in the x- and y-direction. Between these principal Talbot planes ($M/N\in\{0,1/2\}$), additional fractional Talbot planes showing periodic patterns with an increased number of spots can be found. 
Individual Talbot planes ${T_{Q+M/N}}$ are labeled according to their axial position normalized to ${z_T}$ with ${T_0}$ being the original spot array.\\
\figureref{s1} visualizes the generated light field for a \SI{110}{\micro\meter}-pitch microlens array (MLA2, see \tabref{MLAtab1}). As a reference, the intensity distribution behind MLA2 has been computed on basis of the Huygens–Fresnel principle by numerically propagating an incident plane wave. Two xy-sections are presented in \figref[a]{s1}~(left), starting at position of the MLA at $z=\SI{0}{\micro\meter}$ and covering three lenslets in x-direction with the lens centers at $x\in\{\SI{-110}{\micro\meter},\SI{0}{\micro\meter},\SI{110}{\micro\meter}\}$ and $y=\SI{0}{\micro\meter}$. The focal spots of these lenslets are visible in the section for $y=\SI{0}{\micro\meter}$ that intersects the lenslet centers. Their axially nonsymmetric local intensity distribution reflects the expected interplay of refraction and diffraction \cite{Ruffieux2006}. The focal spot array is perfectly recreated at a distance of one Talbot length $z_T = \SI{30.4}{\milli\meter}$. The section at $y=\SI{55}{\micro\meter}$ lies exactly in between two rows of lenslets. Hence, the focal array is absent but the shifted principal plane $T_{1/2}$ is intersected at a distance of $z_T/2$. In the two xz-sections, additional fractional Talbot planes are visible. A closer look at non-principal Talbot planes (e.g.,~$T_{1/4}$) implies a modulation of spot intensities for non-principal Talbot planes \cite{Besold1997a,Kim2015}.\\
With the objective of reducing the computational demands, \figureref[a]{s1}~(right) depicts the two xz-sections for numerical data generated by propagating an array of Gaussian beams matched to the parameters of MLA2. As evident from the plots, the models are in perfect agreement regarding the position of Talbot planes and spot intensities in principal Talbot planes. Therefore, for all further calculations concerning principal Talbot planes the Gaussian beam model has been used, as presented in the main text (\figsref{talbot}, \fref{scaling}) and in this document. Note that even though the plots of \figref{s1} are related to \figref[b]{talbot} of the main text, \figref[b]{talbot} shows vertically projected (i.e.,~integrated) data for the reimaged tweezer array, such that all Talbot planes become visible in one 2D plot.\\
\figureref[b]{s1} displays xy-sections for $T_0$, $T_{1/4}$, $T_{1/2}$. The top row shows calculated data, images in the bottom row are measured intensity distributions (unnormalized). As expected, the focal array $T_0$ is reproduced in the principal plane $T_{1/2}$ while being shifted by $a/2$ in x- and y-direction. For the fractional Talbot plane $T_{1/4}$, the Gaussian beam model reproduces the correct positions and number of spots yet fails to capture the modulation of spot intensities, as anticipated in the discussion of \figref[a]{s1}, which is visible in the imaged intensity distribution. This modulation has also been observed for the atom traps of the similar Talbot plane $T_{-3/4}$ (see \figref[c]{talbot} of the main text).
%
\section{Talbot Tweezer Lattice after reimaging}
\begin{figure}[b]
	\centering
	\includegraphics[width=1\linewidth]{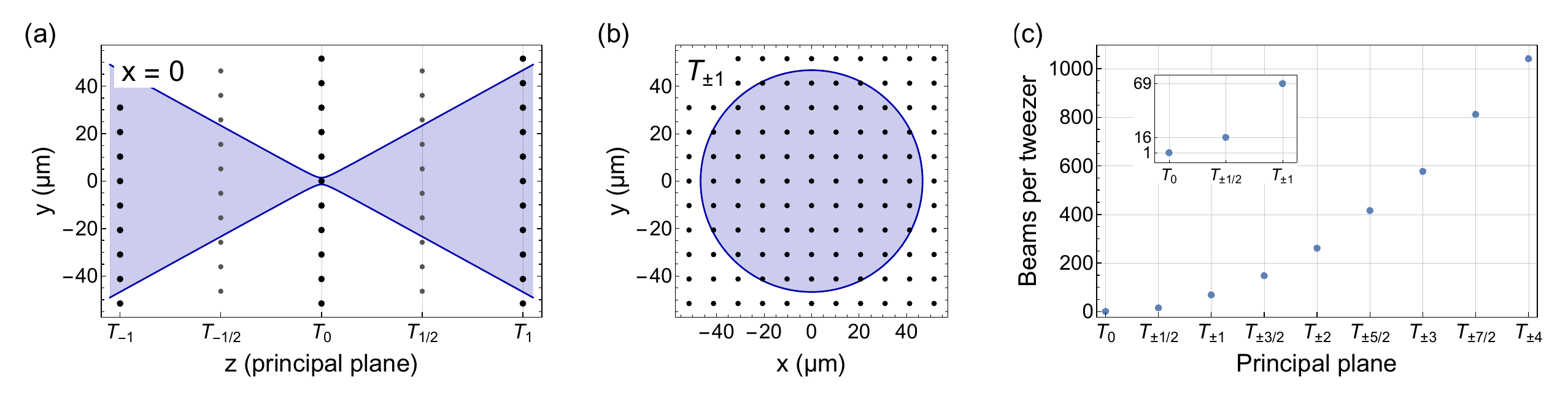}
	\caption{Schematic beam propagation in a Talbot tweezer lattice for the parameters of lattice 2 (see \tabref{MLAtab2}). (a) yz-section along the optical axis and (b) xy-section of $T_{\pm 1}$. The $1/e^2$-waist envelope of the central tweezer is drawn in blue. The Talbot tweezer arrays are depicted by black dots for intersected tweezers. Gray dots visualize the position of xy-shifted principal Talbot planes (e.g.,~$T_{\pm 1/2}$). (c) Scaling of the number of focal array beams (i.e.,~MLA lenslets) contributing to an individual Talbot array tweezer for the different principal planes. This scaling is equivalent to the number of Talbot array tweezers enclosed by the waist envelope of one focal array tweezer as shown in (b).}
	\label{fig:s2}
\end{figure}

\begin{table}[t]%
	\begin{center}%
		\begin{tabular}{p{38mm}p{15mm}p{26mm}p{18mm}p{18mm}p{25mm}p{19mm}}%
			\toprule\addlinespace[6pt]
			\multicolumn{7}{c}{Parameters: Reimaging / Talbot tweezer lattice}\\\addlinespace[2pt]
			\cmidrule(lr){1-7}\\\addlinespace[-8pt]
			Parameter~set\newline\textit{(Figs.\,of main text)}&Microlens array&Geometrical\newline magnification&Reimaged xy-pitch&Talbot length&Radial\,\,(xy)\,\,trap\newline size (measured)&Longitudinal (z) trap size\\
			\midrule
			Lattice~1~\textit{(Figs.\,1,\,3(a),\,4)}&MLA1&375(10)/800(8)&\SI{14.1(4)}{\micro\meter}&\SI{496(28)}{\micro\meter}&\SI{1.45(10)}{\micro\meter}&\SI{8.29(114)}{\micro\meter}\\
			Lattice~2~\textit{(Figs.\,2,\,3)}&MLA2&375(10)/4000(40)&\SI{10.3(3)}{\micro\meter}&\SI{267(15)}{\micro\meter}&\SI{1.45(10)}{\micro\meter}&\SI{8.29(114)}{\micro\meter}\\		
			\bottomrule
		\end{tabular}
	\end{center}
	\caption{Parameters for the implemented Talbot tweezer lattices for a wavelength of $\lambda=\SI{797}{\nano\meter}$. The radial trap size is defined as the minimum beam waist ($1/e^2$ intensity radius) given by the NA of the reimaging optics. The longitudinal trap size is given by the Rayleigh range. For the specified size, tweezers of $U/k_B=\SI{1}{\milli\kelvin}$ depth for $^{85}\text{Rb}$ atoms require \SI{0.6}{\milli\watt} of laser power per tweezer.}
	\label{tab:MLAtab2}
\end{table}%
Experimentally, the Talbot tweezer lattice for atom trapping is realized by reimaging the MLA's focal plane into an apparatus providing laser cooled atoms \cite{Schlosser2012}. Hence, the final parameters are defined by the combination of MLA and relay optics. Typically, the MLA's focal plane is demagnified during reimaging that results in a strong decrease of $a$, now given by the reimaged xy-pitch, and of the Talbot length. In the regime where the Talbot length becomes significantly smaller than the image distance of the optical setup, Talbot planes also appear before the reimaged focal Plane $T_0$. These planes complete the symmetric 3D structure of the resulting Talbot tweezer lattice and are labeled with negative indices. \tableref{MLAtab2} specifies the parameters of the Talbot tweezer lattices implemented in this work. The radial and longitudinal trap size are given by the numerical aperture of the reimaging optics and are thus identical for both implementations.\\
\figureref{s2} gives an estimate of the number of beams (i.e.,~lenslets) contributing to the recreation of the tweezer arrays by the self-imaging effect. In the simple model of \figref[a]{s2}, the waist of the central tweezer of the reimaged focal plane $T_0$ (blue) is drawn in 2D along the optical axis for the parameters of lattice 2 (see \tabref{MLAtab2}). It is evident, that this envelope already encloses a multitude of tweezer sites (gray and black dots) in the closest principal planes. The xy-section for the tweezer array of $T_{\pm 1}$ in \figref[b]{s2} visualizes a number of 69 tweezer sites enclosed by the central tweezer beam of $T_0$. This number equals the number of beams contributing to each tweezer in the Talbot planes $T_{\pm 1}$ \cite{Besold1997b}. The scaling is depicted in \figref[c]{s2} for principal Talbot planes $T_{-4}$ to $T_{4}$. Note that a modification of the tweezer waist would change the beam divergence but not the pitch or Talbot length and thus would directly impact the scaling.
\\
Beyond the simple schematic of \figref{s2}, the finite size of the trapping laser beam on the MLA and therefore the finite  number of lenslets and corresponding focal array tweezers that effectively contribute has to be considered. For the implementation of lattice 2 shown in \figref{scaling} of the main text, the trapping laser beam waist is \SI{1923}{\micro\meter}. This gives a waist of the intensity envelope of the reimaged focal tweezer array $T_0$ of \SI{180}{\micro\meter} and covers an area of 965 tweezers. Experimentally, tweezer depths (i.e.,~maximum intensities) shallower than one fifth of the center tweezer depth are considered to be insufficient to hold atoms, which reduces the effective number of single atom tweezers in $T_0$ to \num{777}. This finite generating pattern causes the decline of tweezer depths in the Talbot tweezer arrays with increasing distance to $T_0$. Propagating a Gaussian beam array with parameters according to the experiment yields 17 relevant principal planes $\{T_0, T_{\pm 1/2}, T_{\pm 1}, T_{\pm 3/2}, T_{\pm 2}, T_{\pm 5/2}, T_{\pm 3}, T_{\pm 7/2}, T_{\pm 4}\}$ containing $\{777, 772, 757, 732, 681, 616, 509, 360, 145\}$ tweezers of sufficient depth (see \figref[c]{scaling} of main text).%
\section{Imperfections and Tolerances}
\begin{figure}%
	\centering%
	\includegraphics[width=1\linewidth]{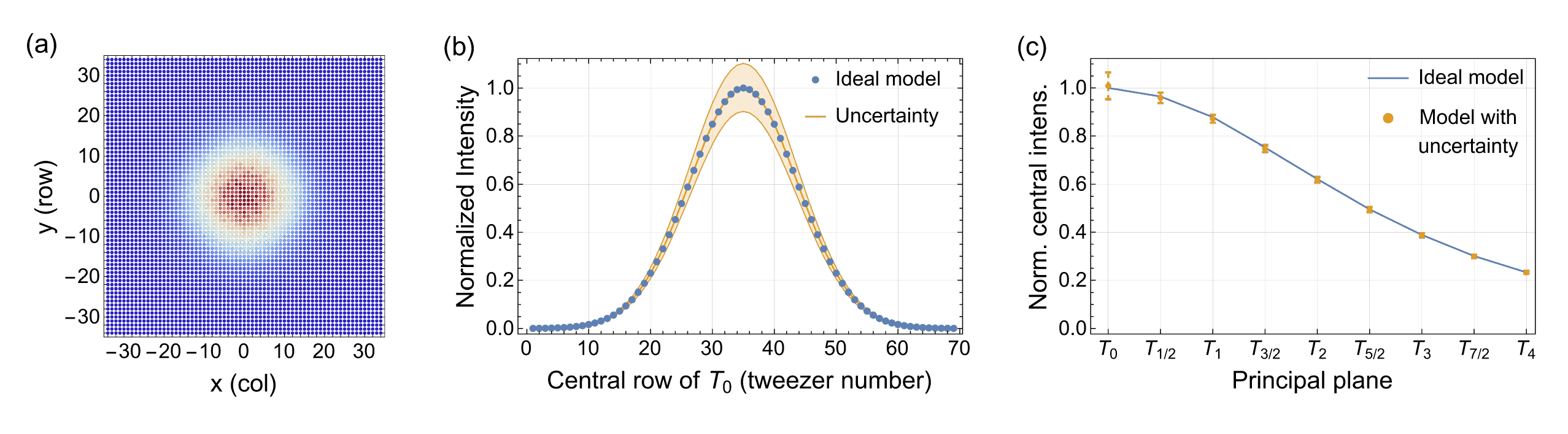}%
	\caption{Talbot tweezer array intensities accounting for uncertainties of lattice 2 and statistical tolerancing. The initial Gaussian intensity profile width is \num{17.5} pitches. For further parameters see Tabs.\,\ref{tab:MLAtab2} and \ref{tab:MLAtab3}. (a) Single realization of $T_0$ tweezer intensities. (b) Ideal tweezer intensities of the central row of $T_0$ (blue dots) and uncertainty (yellow). (c) Central tweezer intensity of principal Talbot planes with increasing distance to $T_0$. The ideal model is shown in blue, data of the tolerance analysis are given in yellow.}%
	\label{fig:s3}%
\end{figure}

\begin{table}[t]%
	\begin{center}
		\begin{tabular}{p{7cm}p{2.5cm}p{3.5cm}}
			\toprule\addlinespace[6pt]
			\multicolumn{3}{c}{Tolerances: Microlens array (MLA2)}\\\addlinespace[2pt]
			\cmidrule(lr){1-3}\\\addlinespace[-8pt]
			Parameter&Value\newline (mean)&Uncertainty\newline (uniform distribution)\\
			\midrule
			Brilliance\newline $\cdot~$estimated E-field amplitude variation&1&\SI{5}{\percent}\\\addlinespace[4pt]
			xy-Pitch\newline $\cdot~$specified value and xy-positioning error&\SI{110}{\micro\meter}&\SI{0,25}{\micro\meter}\\\addlinespace[4pt]
			Radius of curvature\newline $\cdot~$specified value and tolerance&\SI{860}{\micro\meter}&\SI{5}{\percent}\\\addlinespace[4pt]
			Geometric focal length\newline $\cdot~$calculated value and uncertainty of ROC&\SI{1897}{\micro\meter}&\SI{95}{\micro\meter}\\\addlinespace[4pt]
			\toprule\addlinespace[6pt]
			\multicolumn{3}{c}{Uncertainties: Reimaged focal plane / Talbot tweezer array $T_0$}\\\addlinespace[2pt]
			\midrule
			Brilliance\newline $\cdot~$estimated E-field amplitude variation&1&\SI{5}{\percent}\\\addlinespace[4pt]
			xy-Pitch\newline $\cdot~$propagated pitch and xy-positioning error&\SI{10,3125}{\micro\meter}&\SI{0,0234}{\micro\meter}\\\addlinespace[4pt]
			z-Position\newline $\cdot~$propagated focal length error&\num{0} (origin)&\SI{0,83}{\micro\meter}\\\addlinespace[4pt]
			Radial trap size\newline $\cdot~$measured waist ($e^{-2}$ intensity) and uncertainty&\SI{1,45}{\micro\meter}&\SI{0,10}{\micro\meter}\\\addlinespace[4pt]
			Longitudinal trap size\newline $\cdot~$Rayleigh range calculated from waist&\SI{8,29}{\micro\meter}&\SI{1,14}{\micro\meter}\\\addlinespace[4pt]			
			\bottomrule
		\end{tabular}
	\end{center}
	\caption{Parameters ($\lambda=\SI{797}{\nano\meter}$), tolerances of MLA2, and resulting uncertainties for the tweezer array of the reimaged focal plane (Talbot plane $T_0$ of lattice 2).}
	\label{tab:MLAtab3}
\end{table}
A substantial analysis of the real-world performance of Talbot tweezer lattices must include detrimental imperfections of light fields and production tolerances of microlens arrays. As the experiments are conducted with continuous-wave, single-frequency laser light, which is guided by single-mode optical fibers, imperfections in spectral and spatial coherence properties can be neglected (see Ref.\,\cite{Huft2022}, Appendix C). In an approach to capture deviations from a perfect spatial mode (Gaussian mode) of the initial trapping laser beam impinging on the MLA and a variation of the transmittance between individual lenslets, a unified uncertainty of the electromagnetic field amplitude per focal array tweezer of \SI{5}{\percent} has been included. This value is labeled as brilliance in \tabref{MLAtab3}, which also lists tolerances of the MLA and resulting uncertainties for the lattice generating array $T_0$. These have been incorporated in the model of a propagated Gaussian beam array. Global modifications, such as a demagnification error arising from the tolerances of the relay optics and the focus shift caused by the finite aperture of the lenslets have been neglected, as they present a common-mode shift of the mean parameter value but do not add stochastic uncertainties.\\
In order to quantify the consequences for the Talbot tweezer lattice, 400 realizations of lattice 2 generated by an array of 69$\times$69 beams with parameters drawn randomly from a uniform distribution within the uncertainty bounds have been evaluated. \figureref{s3} exemplarily visualizes tweezer intensities. In \figref[a]{s3}, one realization of the intensities for the input beam array is shown. \figureref[b]{s3} depicts the intensities and uncertainty for the central row where the intensity envelope set by the trapping laser beam waist on the MLA corresponds to \num{17.5} pitches. The impact of imperfections in the generating array is illustrated in \figref[c]{s3}. The blue curve gives the central tweezer intensity of all relevant principal planes for an ideal model with zero uncertainty. Yellow dots represent the mean intensity $\bar{I}$ obtained from the tolerance analysis with error bars according to the sample standard deviation:
\begin{align}
	\bar{I}=N^{-1}\sum_{i=1}^N I_i\quad\text{and}\quad\sigma_I=\left[N^{-1}\sum_{i=1}^N(I_i-\bar{I})^2\right]^{1/2}.
\end{align}
\begin{figure}%
	\centering%
	\includegraphics[width=1\linewidth]{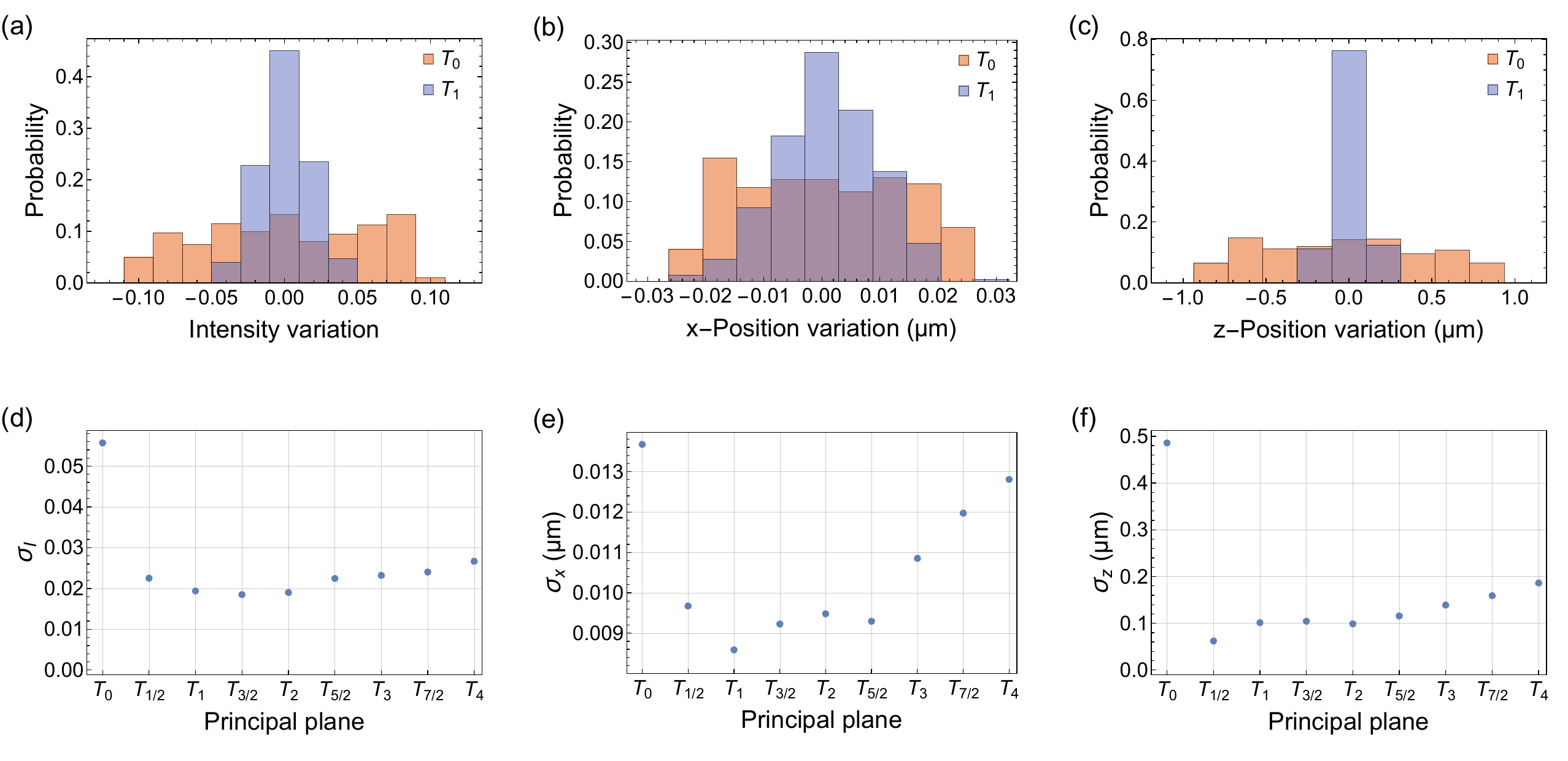}%
	\caption{Statistical tolerance analysis for the central tweezer of the Talbot planes of lattice 2. The analysis is based on 400 random realizations accounting for the uncertainties given in \tabref{MLAtab3}. The top row displays histograms of the variations of (a) the intensity, (b) the x-position and (c) the z-position for the central tweezer of $T_0$ and $T_1$. Variations of the y-position are similar to the variation along x. The bottom row displays the standard deviation of the sample data for these quantities for all relevant Talbot planes.}%
	\label{fig:s4}%
\end{figure}%
As evident from the graph, for all planes the mean tweezer intensity shows negligible deviation (< \SI{1}{\percent}) compared to the ideal model. Furthermore, the standard deviation is strongly reduced for the Talbot planes apart from the generating array ($T_0$). The multi-beam interference creating the self images results in a healing of imperfections, a quality that has been reported for the Talbot effect in other systems as well \cite{Patorski1989}. \figureref{s4} details the statistical tolerances for the central tweezer intensities (a,\,d) as well as their lateral (b,\,e) and axial positions (c,\,f). In the top row, histograms of the variation of these quantities generated from the sample data of $T_0$ and $T_1$ are shown. The bottom row depicts the sample standard deviation for all relevant planes. Again, the striking mitigation of perturbations in the self images is manifested in the data. While the spread of tweezer intensities of $\sigma_I=\num{0.05}$ is more than acceptable for a broad range of applications already and in addition results from a rather ample uncertainty estimate, it can be further improved through preconditioning of the incident trapping laser light using spatial light modulators \cite{Schaffner2020}. Regarding the lateral and axial tweezer positions, statistical tolerances are found to be negligible with a value on the order of single percents of the trap size. Nevertheless, process improvements in MLA manufacturing facilitate further reduction of inhomogeneities \cite{Kirner2019}. As the analysis in \figref{s4} demonstrates, a significant reduction of uncertainties for 2D tweezer arrays can be achieved by the self-stabilizing averaging of trap parameters when operating in principal planes apart from $T_0$. On basis of the presented analysis of light field imperfections and production tolerances of MLAs, Talbot tweezer lattices establish an accessible multilayer platform for quantum science with unprecedented scalability and high optical quality.
\end{document}